\documentclass{sig-alternate}
\usepackage[
  pass,
]{geometry}
\usepackage[utf8]{inputenc}
\usepackage{pgf}
\usepackage[linewidth=1pt]{mdframed}
\usepackage{booktabs}
\usepackage{float}
\usepackage{rotating}
\usepackage{multirow}
\usepackage{graphicx}
\usepackage{color}
\usepackage{xcolor}
\usepackage{subcaption} 
\usepackage{enumitem}
\usepackage{balance}
\usepackage{quoting}

\quotingsetup{vskip=12pt,leftmargin=\leftmargin}

\begin{document}

%
\permission{Copyright is held by the International World Wide Web Conference Committee (IW3C2). IW3C2 reserves the right to provide a hyperlink to the author's site if the Material is used in electronic media.}
\conferenceinfo{WWW 2016,}{April 11--15, 2016, Montr\'eal, Qu\'ebec, Canada.}
\copyrightetc{ACM \the\acmcopyr}
\crdata{978-1-4503-4143-1/16/04. \\
http://dx.doi.org/10.1145/2872427.2883012}

\title{Using Shortlists to Support Decision Making\\and Improve Recommender System Performance}

%
%
%
%
%

\numberofauthors{3} 
%
\author{
%
%
\alignauthor
\hspace{-0.80in}Tobias Schnabel\titlenote{Majority of work performed at Microsoft Research.}\\
\hspace{-0.80in}\affaddr{Cornell University}\\
\hspace{-0.80in}\affaddr{Ithaca, NY, USA}\\
\email{\hspace{-0.75in}tbs49@cornell.edu}
\alignauthor
\hspace*{-0.65in}Paul N.~Bennett, Susan T.~Dumais\\
\hspace*{-0.65in}\affaddr{Microsoft Research}\\
\hspace*{-0.65in}\affaddr{Redmond, WA, USA}\\
\email{\hspace{-0.85in}\{pauben, sdumais\}@microsoft.com\hspace*{-0.35in}}
\alignauthor Thorsten Joachims\\
        \affaddr{Cornell University}\\
        \affaddr{Ithaca, NY, USA}\\
        \email{tj@cs.cornell.edu}
}
\date{1 October 2015}

\maketitle
\begin{abstract}
In this paper, we study {\em shortlists} as an interface component for recommender systems with the dual goal of supporting the user's decision process, as well as improving implicit feedback elicitation for increased recommendation quality. 
A shortlist is a temporary list of candidates that the user is currently considering, e.g., a list of a few  movies the user is currently considering for viewing. From a cognitive perspective, shortlists serve as digital short-term memory where users can off\-load the items under consideration -- thereby decreasing their cognitive load. From a machine learning perspective, adding items to the shortlist generates a new implicit feedback signal as a by-product of exploration and decision making which can improve recommendation quality.  Shortlisting therefore provides additional data for training recommendation systems without the increases in cognitive load that requesting explicit feedback would incur.


We perform an user study with a movie recommendation setup to compare interfaces that offer shortlist support with those that do not.  From the user studies we conclude: (i) users make better decisions with a shortlist; (ii) users prefer an interface with shortlist support; and (iii) the additional implicit feedback from sessions with a shortlist improves the quality of recommendations by nearly a factor of two.

\end{abstract}

 
\vspace*{-0.5\baselineskip} 
\terms{Algorithms, Human Factors, Experimentation}

\vspace*{-0.5\baselineskip} 
\keywords{digital memory, interfaces, decision making, exploration, user engagement, implicit feedback}

\section{Introduction}
Recommender systems play an important role in many online services and websites, including streaming video, music services and e-commerce sites. Within such domains, recommender systems have often succeeded in improving a user's ability to discover desirable items and make informed choices. Designing a successful recommendation system requires careful consideration not only of the machine learning algorithms that underlie the recommendations, but also of the interface through which users interact with the system and generate feedback data. This implies a complex design space of interface usability, incentives to generate data, feedback models, and learning algorithms. 

We explore this design space for the common scenario of decision making and recommendation in \emph{one-choice session-based} tasks. We define these as tasks where (a) users have to make one choice from a large set of options, many of which may be unfamiliar to the user, and (b) the interaction scope is typically limited to one session. Many practical tasks fall into this category, e.g., choosing a movie to watch, comparison shopping, searching for a recipe to make, or picking a hotel. While we assume sessions span only one contiguous chunk of time in this paper, other definitions are possible as long as the context and goal of the user remains the same. For example, a user shopping for a laptop could complete the task over the span of a week, which could be considered a session where the boundaries of a session are task-based rather than time-based. 

\begin{figure*}[t!]%
   \begin{subfigure}[b]{.5\linewidth}
      \centering\includegraphics[width=0.95\textwidth]{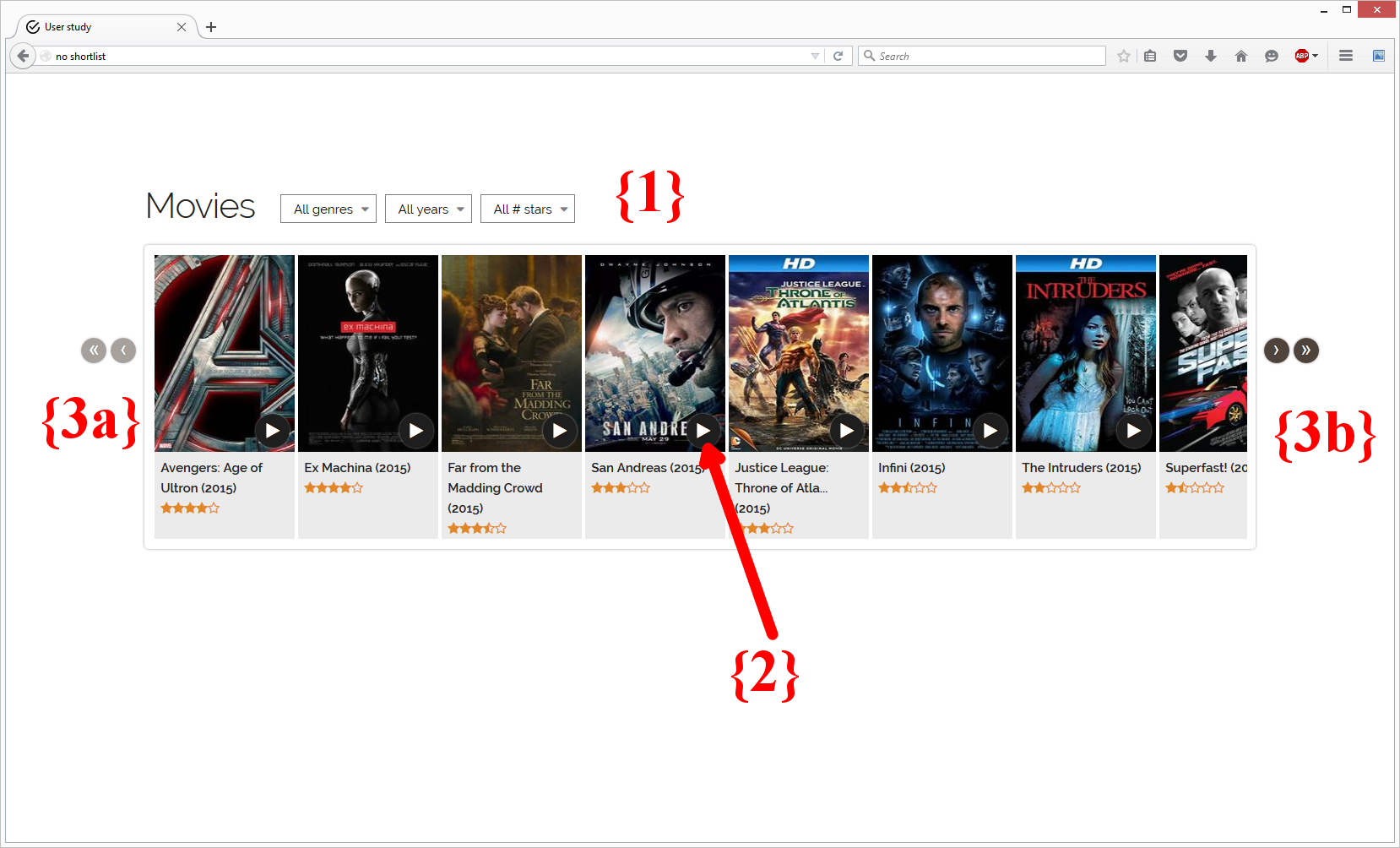}
      \caption{No shortlist}\label{fig:interface_no_short}
   \end{subfigure}%
   \begin{subfigure}[b]{.5\linewidth}
      \centering\includegraphics[width=0.95\textwidth]{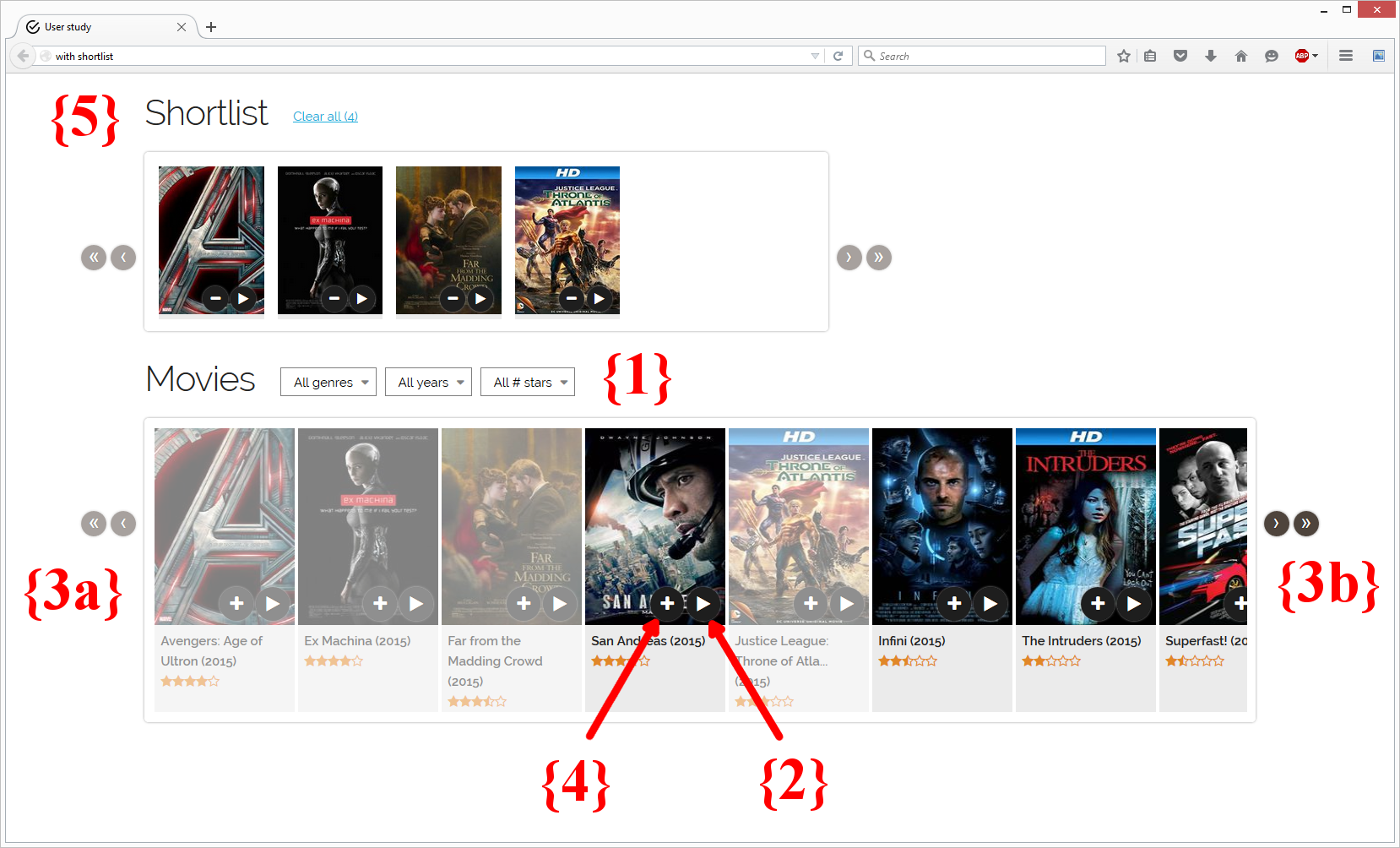}
      \caption{With shortlist}\label{fig:interface_with_short}
   \end{subfigure}
   \caption{The two interfaces each user was presented with.}\label{fig:interfaces}
\end{figure*}

In one-choice session-based tasks there are two important challenges. The first is to provide users with an interface that supports effective decision making by augmenting their cognitive abilities. Assuming a model of bounded rationality \cite{simon1955behavioral}, users are rational agents who want to maximize payoff but under resource constraints. One such resource is short-term memory, and it is well known from research in cognitive psychology that humans have very limited short-term memory \cite{miller1956magical} and that memorizing information incurs a certain cost \cite{baddeley1999essentials}. Interfaces should be designed to alleviate some of these limitations \cite{lidwell2010universal}.

The second challenge is to quickly and non-obtrusively understand a user's needs during a session, such that the recommendation system can provide high-quality recommendations. In order to do this, we need to obtain meaningful and plentiful information about the user's needs during the session. 
This implies that the interface should enable effective feedback collection while minimizing the additional effort that the user has to expend providing feedback. Ideally, a system should encourage implicit feedback as a by-product of their normal information seeking interactions.

In this paper, we propose to improve both usability and feedback elicitation through the introduction of \emph{shortlists}. A shortlist can be thought of as a form of digital memory -- a temporary list of candidates that the user is currently considering. Digital memory is the ability to keep information about the current state, such as interesting items, available in the interface. For example, during a session a user may add a few movies to the shortlist before making a final selection.
Shortlists are different from long-term lists, wish lists, queues or favorites which persist memory across different tasks (i.e., help populate a list of movies that a user plans to watch eventually). We believe that the session-based approach is more appropriate for situations where users make one-time decisions or are heavily influenced by the current context, e.g., when making decisions in a group or on behalf of other people.

This work makes the following three contributions. 
First, we introduce the idea of shortlists with the dual goal of supporting users in their task by enhancing system usability and making the decision process more transparent to the underlying recommender system through the generation of additional implicit feedback. 
Second, we conduct a user study to investigate the impact of the availability of digital memory on the user's exploration behavior, quality of decisions, speed of decision making, cognitive ease of decision making, and overall preference and satisfaction.  
Finally, we investigate whether shortlists lead to user behavior that improves the quality and quantity of implicit user feedback, and whether the use of shortlists therefore leads to better recommendations compared to sessions where the interface had no shortlist. 

Overall, we conclude that shortlists remove cognitive constraints that hinder effective decision making, they improve user satisfaction with the system and the choices users make, and they encourage user behavior that provides valuable implicit feedback to improve recommendation performance.

\section{User Study Design}
In order to study the impact of digital memory on user behavior, we conducted an in-lab user study with a controlled task setup. Among the particular tasks that fall into the category of one-choice session-based tasks, we wanted to pick a task for the study that fulfilled two requirements. First, we wanted a sufficiently large inventory size. This is important since we want to emulate tasks where users are not familiar with all available options -- necessitating exploration; many real-life scenarios are of this nature. Second, we wanted the type of task to be familiar to keep the task instructions to a minimum. The task of selecting a movie to watch from a streaming provider meets these two criteria -- there are a large number of movies to choose from and most people have been exposed to the task as part of their recreational activities. 

\subsection{No Shortlist and Shortlist interfaces}

In our user study, we compare an interface where users were given no digital memory, as represented in Figure~\ref{fig:interface_no_short}, with an interface where session-based digital memory was available via a shortlist, as shown in Figure~\ref{fig:interface_with_short}. 
In both interfaces, users could use facets to filter the current view of movies \{1\} and could use navigation buttons to scroll to the previous and first page \{3a\} (next and last page \{3b\}) of the list.  The facets included drop-downs for year, genre, and review score (on a five star scale). A click on a movie showed more details about the movie such a synopsis of the plot. A click on a movie's play button \{2\} opened a final prompt that asked whether the movie was the user's final selection. 

In the interface with the shortlist (Figure~\ref{fig:interface_with_short}), users could also add movies to a temporary shortlist, \{5\},  while browsing. Users could either drag and drop items from the main list into the shortlist, or click the add button \{4\}. Items within the shortlist could be reordered and also removed. The shortlist interface was inspired by the observation that users often develop strategies for keeping some state in memory. For example, users reported opening multiple tabs in a browser to keep state, or adding items to a shopping cart just to ensure they will be able to remember them. To summarize, both interfaces possessed the same basic functionalities. The only difference is that the shortlist interface in Figure~\ref{fig:interface_with_short} provided the user a way to easily remember and return to items via a shortlist, whereas the interface in Figure~\ref{fig:interface_no_short} possessed no such feature. 

\subsection{Shortlists as session-based memory aids}
At first glance, shortlists might be viewed as another form of shopping carts, or favorite lists. However, shortlists as introduced here are different in two important aspects. First, the purpose of a shortlist is to aid a user in decision making rather than to aid him collect a set of items that he is eventually going to buy. These shortlists simply provide a temporary way of keeping track of items that a user found interesting in the current session, in contrast to favorite lists which express long-term interest. Second, shortlists are visible all the time, making explicit consideration of and comparison with all previously viewed items much easier. In contrast to a simple history of viewed items, shortlists were manually curated and only contained items that users expressed explicit interest in.

\subsection{Study design}
\begin{figure}[t]
    \centering
    \includegraphics[width=\linewidth]{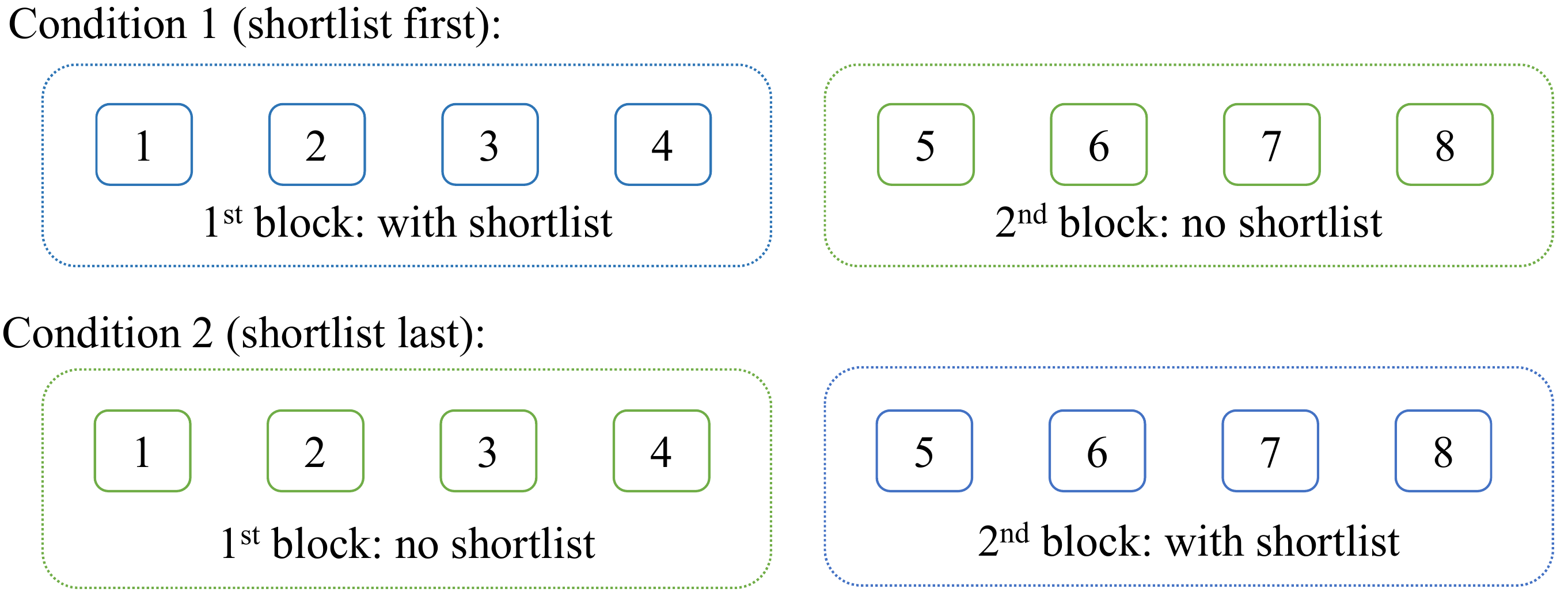}
    \caption{Users were split into two groups starting with different types of blocks.\label{fig:study_layout}\vspace*{-0.75\baselineskip}}
\end{figure}
The overall design is depicted in Figure~\ref{fig:study_layout}. There were two different \emph{blocks}, consisting of four \emph{sessions} each. The type of interface (no shortlist or shortlist) was held constant within a block, but varied across blocks. During each session, users had to pick a movie from a new set of 1030 movies. There was a 3-minute break after the first block, but no further breaks in between sessions.

The sets of movies displayed in each session were disjoint; this was done to prevent a user from learning about available inventory from a previous session. Users were also told that these sets were different. The order and the sets of movies in each session was the same across all users. Users were given the following task statement:
\begin{quotation} 
  \textit{Imagine a very good friend you haven't seen in a year is coming to your place to visit. After hanging out for a while, you plan to watch a movie together. In this experiment, you'll be asked to select a movie to watch with your friend.} \end{quotation}
Users were also asked to keep the same friend in mind for the entire experiment. This prompt was given to emphasize a type of task where session-based preferences play a larger role than long-term preferences. In the future, we would like to study tasks which focus on long-term preferences. We counter-balanced the order of the two conditions across users across conditions to start with a shortlist or not with equal probability. To familiarize users with each interface, before each block we showed a brief video summarizing the main functionality of the interface they would use in the upcoming block.

To summarize, each user performed four repetitions of the same movie selection task with each of the two interfaces for a total of eight sessions per user. Whether the user first experienced the no shortlist or shortlist interface was randomized and balanced across all users.

\subsection{Surveys}
Users completed surveys at the start and the end of the experiment, after each block and also after each session. The pre-experiment questionnaire asked for familiarity with the task and personal investment into the task. After each session, we asked for feedback on the final choice. The surveys after each block asked for the immediate experience with the interface and for self-reported strategies and goals. In the final questionnaire, we asked users to compare both interfaces and for their overall preferences.  Interfaces were referred to as ``first'' and ``second'' interface with an illustration similar to Figure \ref{fig:interfaces} to avoid framing biases from the wording. After the entire experiment, we debriefed users in a short oral session and asked them for any other comments they had on the experiment.

\subsection{Data}
We obtained the movie data from OMDb\footnote{http://omdbapi.com/}, a free and open movie database. We only selected movies that appeared in the year 1980 or after and with sufficiently many votes on IMDb (800 or more). This filtering step was done to ensure all movies had a general level of attractiveness and popularity. We partitioned this set of movies into eight (non-overlapping) subsets of size 1030 each (for 8240 movies total). These partitions were held constant across users in the study. The order displayed to the user was first descending by year and then descending by IMDb score.  This was to ensure the no shortlist condition offered a reasonable baseline for the condition.  Note that with this default ordering users see recent highly-rated movies first.

\subsection{Users}
We recruited 60 people for the user study; most were graduate students in STEM fields. There were 15 female and 45 male participants, yielding a gender ratio of 25\% to 75\%. All users were given the same computer and monitor to avoid differences in hardware affecting user behavior.

\section{User Study Results}
\label{sec:user-study}
In this section, we address the impact of shortlists in terms of user outcome. The goal of our user study was to answer the following research questions: 
\begin{enumerate}
\item How is overall user satisfaction changed by the ability to shortlist? When comparing an interface with a shortlist against an interface without, which one would users prefer?
\item How does the shortlist influence the perceived quality of decisions? 
\item How do shortlists alter exploration? More specifically, do shortlists influence the time-to-decision or the number of items that were explored before making a decision?
\end{enumerate}

For the remainder of the paper, we will employ the following terminology.
We refer to an item as \emph{displayed} if an item was visible on the users viewport. In other words, this means an item was shown in the main list at some point during the session. For example, there are eight displayed items in Figure~\ref{fig:interface_no_short}.
An \emph{examined} item is an item which got clicked on by the user in order to open a detail page with more information such as a plot synopsis, reviews, etc.
\emph{Shortlisted} items are items that the users added to the shortlist at some point during the session. As an example, there are four shortlisted items in Figure~\ref{fig:interface_with_short}.
Finally, a \emph{chosen} item is the one item in a session that the user picked as his final choice. In total, each user had to choose eight movies, where four of them were chosen with shortlist support, and four of them were chosen without shortlist support.
The following two subsections will present results for the user interface aspects -- how do people like shortlists as interface components? After that, we turn to the behavioral aspects, showing that shortlists do indeed change exploration and decision strategies. 

\subsection{People prefer and use shortlists}
\begin{figure}[t]
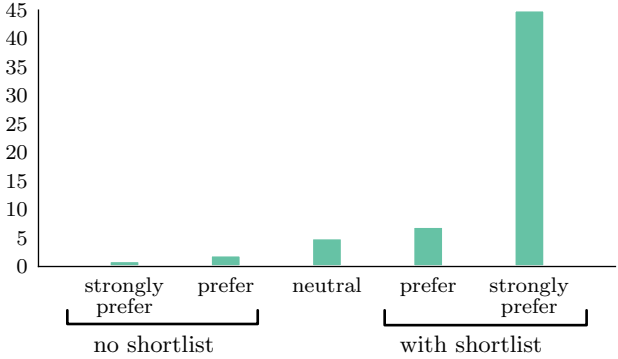

    \centering
\begingroup%
\makeatletter%
\begin{pgfpicture}%
\pgfpathrectangle{\pgfpointorigin}{\pgfqpoint{3.320880in}{2.052417in}}%
\pgfusepath{use as bounding box, clip}%
\begin{pgfscope}%
\pgfsetbuttcap%
\pgfsetmiterjoin%
\definecolor{currentfill}{rgb}{1.000000,1.000000,1.000000}%
\pgfsetfillcolor{currentfill}%
\pgfsetlinewidth{0.000000pt}%
\definecolor{currentstroke}{rgb}{1.000000,1.000000,1.000000}%
\pgfsetstrokecolor{currentstroke}%
\pgfsetdash{}{0pt}%
\pgfpathmoveto{\pgfqpoint{0.000000in}{0.000000in}}%
\pgfpathlineto{\pgfqpoint{3.320880in}{0.000000in}}%
\pgfpathlineto{\pgfqpoint{3.320880in}{2.052417in}}%
\pgfpathlineto{\pgfqpoint{0.000000in}{2.052417in}}%
\pgfpathclose%
\pgfusepath{fill}%
\end{pgfscope}%
\begin{pgfscope}%
\pgfsetbuttcap%
\pgfsetmiterjoin%
\definecolor{currentfill}{rgb}{1.000000,1.000000,1.000000}%
\pgfsetfillcolor{currentfill}%
\pgfsetlinewidth{0.000000pt}%
\definecolor{currentstroke}{rgb}{0.000000,0.000000,0.000000}%
\pgfsetstrokecolor{currentstroke}%
\pgfsetstrokeopacity{0.000000}%
\pgfsetdash{}{0pt}%
\pgfpathmoveto{\pgfqpoint{0.166044in}{0.513104in}}%
\pgfpathlineto{\pgfqpoint{3.185880in}{0.513104in}}%
\pgfpathlineto{\pgfqpoint{3.185880in}{1.856089in}}%
\pgfpathlineto{\pgfqpoint{0.166044in}{1.856089in}}%
\pgfpathclose%
\pgfusepath{fill}%
\end{pgfscope}%
\begin{pgfscope}%
\pgfpathrectangle{\pgfqpoint{0.166044in}{0.513104in}}{\pgfqpoint{3.019836in}{1.342985in}} %
\pgfusepath{clip}%
\pgfsetbuttcap%
\pgfsetmiterjoin%
\definecolor{currentfill}{rgb}{0.400000,0.760784,0.647059}%
\pgfsetfillcolor{currentfill}%
\pgfsetlinewidth{1.003750pt}%
\definecolor{currentstroke}{rgb}{1.000000,1.000000,1.000000}%
\pgfsetstrokecolor{currentstroke}%
\pgfsetdash{}{0pt}%
\pgfpathmoveto{\pgfqpoint{0.536901in}{0.513104in}}%
\pgfpathlineto{\pgfqpoint{0.695840in}{0.513104in}}%
\pgfpathlineto{\pgfqpoint{0.695840in}{0.542948in}}%
\pgfpathlineto{\pgfqpoint{0.536901in}{0.542948in}}%
\pgfpathlineto{\pgfqpoint{0.536901in}{0.513104in}}%
\pgfusepath{stroke,fill}%
\end{pgfscope}%
\begin{pgfscope}%
\pgfpathrectangle{\pgfqpoint{0.166044in}{0.513104in}}{\pgfqpoint{3.019836in}{1.342985in}} %
\pgfusepath{clip}%
\pgfsetbuttcap%
\pgfsetmiterjoin%
\definecolor{currentfill}{rgb}{0.400000,0.760784,0.647059}%
\pgfsetfillcolor{currentfill}%
\pgfsetlinewidth{1.003750pt}%
\definecolor{currentstroke}{rgb}{1.000000,1.000000,1.000000}%
\pgfsetstrokecolor{currentstroke}%
\pgfsetdash{}{0pt}%
\pgfpathmoveto{\pgfqpoint{1.066697in}{0.513104in}}%
\pgfpathlineto{\pgfqpoint{1.225636in}{0.513104in}}%
\pgfpathlineto{\pgfqpoint{1.225636in}{0.572792in}}%
\pgfpathlineto{\pgfqpoint{1.066697in}{0.572792in}}%
\pgfpathlineto{\pgfqpoint{1.066697in}{0.513104in}}%
\pgfusepath{stroke,fill}%
\end{pgfscope}%
\begin{pgfscope}%
\pgfpathrectangle{\pgfqpoint{0.166044in}{0.513104in}}{\pgfqpoint{3.019836in}{1.342985in}} %
\pgfusepath{clip}%
\pgfsetbuttcap%
\pgfsetmiterjoin%
\definecolor{currentfill}{rgb}{0.400000,0.760784,0.647059}%
\pgfsetfillcolor{currentfill}%
\pgfsetlinewidth{1.003750pt}%
\definecolor{currentstroke}{rgb}{1.000000,1.000000,1.000000}%
\pgfsetstrokecolor{currentstroke}%
\pgfsetdash{}{0pt}%
\pgfpathmoveto{\pgfqpoint{1.596493in}{0.513104in}}%
\pgfpathlineto{\pgfqpoint{1.755431in}{0.513104in}}%
\pgfpathlineto{\pgfqpoint{1.755431in}{0.662325in}}%
\pgfpathlineto{\pgfqpoint{1.596493in}{0.662325in}}%
\pgfpathlineto{\pgfqpoint{1.596493in}{0.513104in}}%
\pgfusepath{stroke,fill}%
\end{pgfscope}%
\begin{pgfscope}%
\pgfpathrectangle{\pgfqpoint{0.166044in}{0.513104in}}{\pgfqpoint{3.019836in}{1.342985in}} %
\pgfusepath{clip}%
\pgfsetbuttcap%
\pgfsetmiterjoin%
\definecolor{currentfill}{rgb}{0.400000,0.760784,0.647059}%
\pgfsetfillcolor{currentfill}%
\pgfsetlinewidth{1.003750pt}%
\definecolor{currentstroke}{rgb}{1.000000,1.000000,1.000000}%
\pgfsetstrokecolor{currentstroke}%
\pgfsetdash{}{0pt}%
\pgfpathmoveto{\pgfqpoint{2.126288in}{0.513104in}}%
\pgfpathlineto{\pgfqpoint{2.285227in}{0.513104in}}%
\pgfpathlineto{\pgfqpoint{2.285227in}{0.722013in}}%
\pgfpathlineto{\pgfqpoint{2.126288in}{0.722013in}}%
\pgfpathlineto{\pgfqpoint{2.126288in}{0.513104in}}%
\pgfusepath{stroke,fill}%
\end{pgfscope}%
\begin{pgfscope}%
\pgfpathrectangle{\pgfqpoint{0.166044in}{0.513104in}}{\pgfqpoint{3.019836in}{1.342985in}} %
\pgfusepath{clip}%
\pgfsetbuttcap%
\pgfsetmiterjoin%
\definecolor{currentfill}{rgb}{0.400000,0.760784,0.647059}%
\pgfsetfillcolor{currentfill}%
\pgfsetlinewidth{1.003750pt}%
\definecolor{currentstroke}{rgb}{1.000000,1.000000,1.000000}%
\pgfsetstrokecolor{currentstroke}%
\pgfsetdash{}{0pt}%
\pgfpathmoveto{\pgfqpoint{2.656084in}{0.513104in}}%
\pgfpathlineto{\pgfqpoint{2.815023in}{0.513104in}}%
\pgfpathlineto{\pgfqpoint{2.815023in}{1.856089in}}%
\pgfpathlineto{\pgfqpoint{2.656084in}{1.856089in}}%
\pgfpathlineto{\pgfqpoint{2.656084in}{0.513104in}}%
\pgfusepath{stroke,fill}%
\end{pgfscope}%
\begin{pgfscope}%
\pgfsetrectcap%
\pgfsetmiterjoin%
\pgfsetlinewidth{0.501875pt}%
\definecolor{currentstroke}{rgb}{0.000000,0.000000,0.000000}%
\pgfsetstrokecolor{currentstroke}%
\pgfsetdash{}{0pt}%
\pgfpathmoveto{\pgfqpoint{0.166044in}{0.513104in}}%
\pgfpathlineto{\pgfqpoint{3.185880in}{0.513104in}}%
\pgfusepath{stroke}%
\end{pgfscope}%
\begin{pgfscope}%
\pgfsetrectcap%
\pgfsetmiterjoin%
\pgfsetlinewidth{0.501875pt}%
\definecolor{currentstroke}{rgb}{0.000000,0.000000,0.000000}%
\pgfsetstrokecolor{currentstroke}%
\pgfsetdash{}{0pt}%
\pgfpathmoveto{\pgfqpoint{0.166044in}{0.513104in}}%
\pgfpathlineto{\pgfqpoint{0.166044in}{1.856089in}}%
\pgfusepath{stroke}%
\end{pgfscope}%
\begin{pgfscope}%
\pgftext[x=0.407808in,y=0.381024in,left,base]{\rmfamily\fontsize{8.000000}{9.600000}\selectfont strongly}%
\end{pgfscope}%
\begin{pgfscope}%
\pgftext[x=0.467101in,y=0.267595in,left,base]{\rmfamily\fontsize{8.000000}{9.600000}\selectfont prefer}%
\end{pgfscope}%
\begin{pgfscope}%
\pgftext[x=1.146166in,y=0.457549in,,top]{\rmfamily\fontsize{8.000000}{9.600000}\selectfont prefer}%
\end{pgfscope}%
\begin{pgfscope}%
\pgftext[x=1.675962in,y=0.457549in,,top]{\rmfamily\fontsize{8.000000}{9.600000}\selectfont neutral}%
\end{pgfscope}%
\begin{pgfscope}%
\pgftext[x=2.205758in,y=0.457549in,,top]{\rmfamily\fontsize{8.000000}{9.600000}\selectfont prefer}%
\end{pgfscope}%
\begin{pgfscope}%
\pgftext[x=2.526992in,y=0.381024in,left,base]{\rmfamily\fontsize{8.000000}{9.600000}\selectfont strongly}%
\end{pgfscope}%
\begin{pgfscope}%
\pgftext[x=2.586285in,y=0.267595in,left,base]{\rmfamily\fontsize{8.000000}{9.600000}\selectfont prefer}%
\end{pgfscope}%
\begin{pgfscope}%
\pgftext[x=0.110488in,y=0.513104in,right,]{\rmfamily\fontsize{8.000000}{9.600000}\selectfont \(\displaystyle 0\)}%
\end{pgfscope}%
\begin{pgfscope}%
\pgftext[x=0.110488in,y=0.662325in,right,]{\rmfamily\fontsize{8.000000}{9.600000}\selectfont \(\displaystyle 5\)}%
\end{pgfscope}%
\begin{pgfscope}%
\pgftext[x=0.110488in,y=0.811545in,right,]{\rmfamily\fontsize{8.000000}{9.600000}\selectfont \(\displaystyle 10\)}%
\end{pgfscope}%
\begin{pgfscope}%
\pgftext[x=0.110488in,y=0.960766in,right,]{\rmfamily\fontsize{8.000000}{9.600000}\selectfont \(\displaystyle 15\)}%
\end{pgfscope}%
\begin{pgfscope}%
\pgftext[x=0.110488in,y=1.109986in,right,]{\rmfamily\fontsize{8.000000}{9.600000}\selectfont \(\displaystyle 20\)}%
\end{pgfscope}%
\begin{pgfscope}%
\pgftext[x=0.110488in,y=1.259207in,right,]{\rmfamily\fontsize{8.000000}{9.600000}\selectfont \(\displaystyle 25\)}%
\end{pgfscope}%
\begin{pgfscope}%
\pgftext[x=0.110488in,y=1.408427in,right,]{\rmfamily\fontsize{8.000000}{9.600000}\selectfont \(\displaystyle 30\)}%
\end{pgfscope}%
\begin{pgfscope}%
\pgftext[x=0.110488in,y=1.557648in,right,]{\rmfamily\fontsize{8.000000}{9.600000}\selectfont \(\displaystyle 35\)}%
\end{pgfscope}%
\begin{pgfscope}%
\pgftext[x=0.110488in,y=1.706869in,right,]{\rmfamily\fontsize{8.000000}{9.600000}\selectfont \(\displaystyle 40\)}%
\end{pgfscope}%
\begin{pgfscope}%
\pgftext[x=0.110488in,y=1.856089in,right,]{\rmfamily\fontsize{8.000000}{9.600000}\selectfont \(\displaystyle 45\)}%
\end{pgfscope}%
\begin{pgfscope}%
\pgfsetroundcap%
\pgfsetroundjoin%
\pgfsetlinewidth{1.003750pt}%
\definecolor{currentstroke}{rgb}{0.000000,0.000000,0.000000}%
\pgfsetstrokecolor{currentstroke}%
\pgfsetdash{}{0pt}%
\pgfpathmoveto{\pgfqpoint{1.313582in}{0.283872in}}%
\pgfpathlineto{\pgfqpoint{1.313582in}{0.212002in}}%
\pgfpathlineto{\pgfqpoint{0.317036in}{0.212002in}}%
\pgfpathlineto{\pgfqpoint{0.317036in}{0.283872in}}%
\pgfusepath{stroke}%
\end{pgfscope}%
\begin{pgfscope}%
\pgftext[x=0.770011in,y=0.110209in,,]{\rmfamily\fontsize{9.000000}{10.800000}\selectfont no shortlist}%
\end{pgfscope}%
\begin{pgfscope}%
\pgfsetroundcap%
\pgfsetroundjoin%
\pgfsetlinewidth{1.003750pt}%
\definecolor{currentstroke}{rgb}{0.000000,0.000000,0.000000}%
\pgfsetstrokecolor{currentstroke}%
\pgfsetdash{}{0pt}%
\pgfpathmoveto{\pgfqpoint{3.034888in}{0.283865in}}%
\pgfpathlineto{\pgfqpoint{3.034888in}{0.205962in}}%
\pgfpathlineto{\pgfqpoint{1.977946in}{0.205962in}}%
\pgfpathlineto{\pgfqpoint{1.977946in}{0.283865in}}%
\pgfusepath{stroke}%
\end{pgfscope}%
\begin{pgfscope}%
\pgftext[x=2.430921in,y=0.110209in,,]{\rmfamily\fontsize{9.000000}{10.800000}\selectfont with shortlist}%
\end{pgfscope}%
\end{pgfpicture}%
\makeatother%
\endgroup%
    \caption{Most users prefer the interface with the shortlist.\label{fig:preferred_interface}}
\end{figure}

One of the most important aspects for long-term engagement with a system is user satisfaction. At the end of the experiment, we asked users to indicate their relative preference with respect to the two interfaces on a five-point scale. As mentioned earlier, the interfaces were referred to as ``first interface" and ``second interface" to avoid framing biases from the wording.

The results of this question are displayed in Figure~\ref{fig:preferred_interface}. The vast majority of users (52) either prefers or strongly prefers the interface with the shortlist ($p < 0.001$; binomial test). This is also in line with what users entered in the feedback section that allowed for free form text or told us during the debriefing session. 
Users reported decreased cognitive load when they could save interesting items in the shortlist. Another popular comment was that the shortlist helped users in their task by being able to compare items directly. 

We can also see the overall user satisfaction reflected in the number of times that they actually interact with the shortlist. Of the 240 sessions where users had the shortlist interface (four sessions per user, 60 users in total), people used shortlists in 224 cases. In other words, people used shortlists in over 93\% of the sessions where it was available. This is despite the fact that using the shortlist was optional, and at no point in the study did we ask them to use any particular function of the user interface. The high repeat usage of the interface also indicates that there is repeated benefits that users get out of the shortlist.

\begin{table}[htb]
  \centering
    \begin{tabular}{rr}
    \toprule
    \# sessions shortlist used & users \\
    \midrule
    1     & 1 \\
    2     & 4 \\
    3     & 5 \\
    4     & 50 \\
    \bottomrule
    \end{tabular}
    \caption{Most users employed shortlists in all sessions.\label{tbl:shortlist_usage}}
\end{table}


It is also interesting to look at the distribution of users with respect to shortlist usage.  Table~\ref{tbl:shortlist_usage} reports the number of users grouped by the number of sessions in which they used shortlists. The first observation we make is that over 80\% of the users used shortlisting in all four sessions where it was available. Also, every user tried the shortlist interface, although one user tried it in only one session. In summary, we conclude shortlists have a high task-related affordance as indicated by the high and consistent usage throughout sessions and the overall preference for the shortlist interface.

\subsection{Higher choice satisfaction with shortlists}
As we saw in the previous subsection, users prefer the shortlisting interface over a regular interface. A natural question is whether that also translates to the choices they make using the interfaces. In order to answer this, we elicited responses asking users to self-assess overall and per-session satisfaction.  For overall, we asked users the following question in the final survey: ``In which interface were you most satisfied with your selections?". We also asked directly after each session for an absolute judgement on a five-point scale (1-5) of how satisfied users were with their current selection.

\begin{figure}[t]
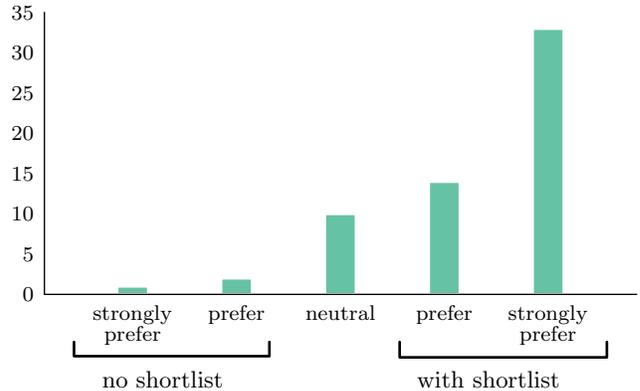

    \centering
\begingroup%
\makeatletter%
\begin{pgfpicture}%
\pgfpathrectangle{\pgfpointorigin}{\pgfqpoint{3.320880in}{2.052417in}}%
\pgfusepath{use as bounding box, clip}%
\begin{pgfscope}%
\pgfsetbuttcap%
\pgfsetmiterjoin%
\definecolor{currentfill}{rgb}{1.000000,1.000000,1.000000}%
\pgfsetfillcolor{currentfill}%
\pgfsetlinewidth{0.000000pt}%
\definecolor{currentstroke}{rgb}{1.000000,1.000000,1.000000}%
\pgfsetstrokecolor{currentstroke}%
\pgfsetdash{}{0pt}%
\pgfpathmoveto{\pgfqpoint{0.000000in}{0.000000in}}%
\pgfpathlineto{\pgfqpoint{3.320880in}{0.000000in}}%
\pgfpathlineto{\pgfqpoint{3.320880in}{2.052417in}}%
\pgfpathlineto{\pgfqpoint{0.000000in}{2.052417in}}%
\pgfpathclose%
\pgfusepath{fill}%
\end{pgfscope}%
\begin{pgfscope}%
\pgfsetbuttcap%
\pgfsetmiterjoin%
\definecolor{currentfill}{rgb}{1.000000,1.000000,1.000000}%
\pgfsetfillcolor{currentfill}%
\pgfsetlinewidth{0.000000pt}%
\definecolor{currentstroke}{rgb}{0.000000,0.000000,0.000000}%
\pgfsetstrokecolor{currentstroke}%
\pgfsetstrokeopacity{0.000000}%
\pgfsetdash{}{0pt}%
\pgfpathmoveto{\pgfqpoint{0.166044in}{0.513104in}}%
\pgfpathlineto{\pgfqpoint{3.264840in}{0.513104in}}%
\pgfpathlineto{\pgfqpoint{3.264840in}{1.989455in}}%
\pgfpathlineto{\pgfqpoint{0.166044in}{1.989455in}}%
\pgfpathclose%
\pgfusepath{fill}%
\end{pgfscope}%
\begin{pgfscope}%
\pgfpathrectangle{\pgfqpoint{0.166044in}{0.513104in}}{\pgfqpoint{3.098796in}{1.476351in}} %
\pgfusepath{clip}%
\pgfsetbuttcap%
\pgfsetmiterjoin%
\definecolor{currentfill}{rgb}{0.400000,0.760784,0.647059}%
\pgfsetfillcolor{currentfill}%
\pgfsetlinewidth{1.003750pt}%
\definecolor{currentstroke}{rgb}{1.000000,1.000000,1.000000}%
\pgfsetstrokecolor{currentstroke}%
\pgfsetdash{}{0pt}%
\pgfpathmoveto{\pgfqpoint{0.546598in}{0.513104in}}%
\pgfpathlineto{\pgfqpoint{0.709692in}{0.513104in}}%
\pgfpathlineto{\pgfqpoint{0.709692in}{0.555286in}}%
\pgfpathlineto{\pgfqpoint{0.546598in}{0.555286in}}%
\pgfpathlineto{\pgfqpoint{0.546598in}{0.513104in}}%
\pgfusepath{stroke,fill}%
\end{pgfscope}%
\begin{pgfscope}%
\pgfpathrectangle{\pgfqpoint{0.166044in}{0.513104in}}{\pgfqpoint{3.098796in}{1.476351in}} %
\pgfusepath{clip}%
\pgfsetbuttcap%
\pgfsetmiterjoin%
\definecolor{currentfill}{rgb}{0.400000,0.760784,0.647059}%
\pgfsetfillcolor{currentfill}%
\pgfsetlinewidth{1.003750pt}%
\definecolor{currentstroke}{rgb}{1.000000,1.000000,1.000000}%
\pgfsetstrokecolor{currentstroke}%
\pgfsetdash{}{0pt}%
\pgfpathmoveto{\pgfqpoint{1.090246in}{0.513104in}}%
\pgfpathlineto{\pgfqpoint{1.253341in}{0.513104in}}%
\pgfpathlineto{\pgfqpoint{1.253341in}{0.597467in}}%
\pgfpathlineto{\pgfqpoint{1.090246in}{0.597467in}}%
\pgfpathlineto{\pgfqpoint{1.090246in}{0.513104in}}%
\pgfusepath{stroke,fill}%
\end{pgfscope}%
\begin{pgfscope}%
\pgfpathrectangle{\pgfqpoint{0.166044in}{0.513104in}}{\pgfqpoint{3.098796in}{1.476351in}} %
\pgfusepath{clip}%
\pgfsetbuttcap%
\pgfsetmiterjoin%
\definecolor{currentfill}{rgb}{0.400000,0.760784,0.647059}%
\pgfsetfillcolor{currentfill}%
\pgfsetlinewidth{1.003750pt}%
\definecolor{currentstroke}{rgb}{1.000000,1.000000,1.000000}%
\pgfsetstrokecolor{currentstroke}%
\pgfsetdash{}{0pt}%
\pgfpathmoveto{\pgfqpoint{1.633895in}{0.513104in}}%
\pgfpathlineto{\pgfqpoint{1.796989in}{0.513104in}}%
\pgfpathlineto{\pgfqpoint{1.796989in}{0.934919in}}%
\pgfpathlineto{\pgfqpoint{1.633895in}{0.934919in}}%
\pgfpathlineto{\pgfqpoint{1.633895in}{0.513104in}}%
\pgfusepath{stroke,fill}%
\end{pgfscope}%
\begin{pgfscope}%
\pgfpathrectangle{\pgfqpoint{0.166044in}{0.513104in}}{\pgfqpoint{3.098796in}{1.476351in}} %
\pgfusepath{clip}%
\pgfsetbuttcap%
\pgfsetmiterjoin%
\definecolor{currentfill}{rgb}{0.400000,0.760784,0.647059}%
\pgfsetfillcolor{currentfill}%
\pgfsetlinewidth{1.003750pt}%
\definecolor{currentstroke}{rgb}{1.000000,1.000000,1.000000}%
\pgfsetstrokecolor{currentstroke}%
\pgfsetdash{}{0pt}%
\pgfpathmoveto{\pgfqpoint{2.177543in}{0.513104in}}%
\pgfpathlineto{\pgfqpoint{2.340638in}{0.513104in}}%
\pgfpathlineto{\pgfqpoint{2.340638in}{1.103644in}}%
\pgfpathlineto{\pgfqpoint{2.177543in}{1.103644in}}%
\pgfpathlineto{\pgfqpoint{2.177543in}{0.513104in}}%
\pgfusepath{stroke,fill}%
\end{pgfscope}%
\begin{pgfscope}%
\pgfpathrectangle{\pgfqpoint{0.166044in}{0.513104in}}{\pgfqpoint{3.098796in}{1.476351in}} %
\pgfusepath{clip}%
\pgfsetbuttcap%
\pgfsetmiterjoin%
\definecolor{currentfill}{rgb}{0.400000,0.760784,0.647059}%
\pgfsetfillcolor{currentfill}%
\pgfsetlinewidth{1.003750pt}%
\definecolor{currentstroke}{rgb}{1.000000,1.000000,1.000000}%
\pgfsetstrokecolor{currentstroke}%
\pgfsetdash{}{0pt}%
\pgfpathmoveto{\pgfqpoint{2.721192in}{0.513104in}}%
\pgfpathlineto{\pgfqpoint{2.884286in}{0.513104in}}%
\pgfpathlineto{\pgfqpoint{2.884286in}{1.905092in}}%
\pgfpathlineto{\pgfqpoint{2.721192in}{1.905092in}}%
\pgfpathlineto{\pgfqpoint{2.721192in}{0.513104in}}%
\pgfusepath{stroke,fill}%
\end{pgfscope}%
\begin{pgfscope}%
\pgfsetrectcap%
\pgfsetmiterjoin%
\pgfsetlinewidth{0.501875pt}%
\definecolor{currentstroke}{rgb}{0.000000,0.000000,0.000000}%
\pgfsetstrokecolor{currentstroke}%
\pgfsetdash{}{0pt}%
\pgfpathmoveto{\pgfqpoint{0.166044in}{0.513104in}}%
\pgfpathlineto{\pgfqpoint{3.264840in}{0.513104in}}%
\pgfusepath{stroke}%
\end{pgfscope}%
\begin{pgfscope}%
\pgfsetrectcap%
\pgfsetmiterjoin%
\pgfsetlinewidth{0.501875pt}%
\definecolor{currentstroke}{rgb}{0.000000,0.000000,0.000000}%
\pgfsetstrokecolor{currentstroke}%
\pgfsetdash{}{0pt}%
\pgfpathmoveto{\pgfqpoint{0.166044in}{0.513104in}}%
\pgfpathlineto{\pgfqpoint{0.166044in}{1.989455in}}%
\pgfusepath{stroke}%
\end{pgfscope}%
\begin{pgfscope}%
\pgftext[x=0.419583in,y=0.381024in,left,base]{\rmfamily\fontsize{8.000000}{9.600000}\selectfont strongly}%
\end{pgfscope}%
\begin{pgfscope}%
\pgftext[x=0.478876in,y=0.267595in,left,base]{\rmfamily\fontsize{8.000000}{9.600000}\selectfont prefer}%
\end{pgfscope}%
\begin{pgfscope}%
\pgftext[x=1.171794in,y=0.457549in,,top]{\rmfamily\fontsize{8.000000}{9.600000}\selectfont prefer}%
\end{pgfscope}%
\begin{pgfscope}%
\pgftext[x=1.715442in,y=0.457549in,,top]{\rmfamily\fontsize{8.000000}{9.600000}\selectfont neutral}%
\end{pgfscope}%
\begin{pgfscope}%
\pgftext[x=2.259091in,y=0.457549in,,top]{\rmfamily\fontsize{8.000000}{9.600000}\selectfont prefer}%
\end{pgfscope}%
\begin{pgfscope}%
\pgftext[x=2.594177in,y=0.381024in,left,base]{\rmfamily\fontsize{8.000000}{9.600000}\selectfont strongly}%
\end{pgfscope}%
\begin{pgfscope}%
\pgftext[x=2.653470in,y=0.267595in,left,base]{\rmfamily\fontsize{8.000000}{9.600000}\selectfont prefer}%
\end{pgfscope}%
\begin{pgfscope}%
\pgftext[x=0.110488in,y=0.513104in,right,]{\rmfamily\fontsize{8.000000}{9.600000}\selectfont \(\displaystyle 0\)}%
\end{pgfscope}%
\begin{pgfscope}%
\pgftext[x=0.110488in,y=0.724011in,right,]{\rmfamily\fontsize{8.000000}{9.600000}\selectfont \(\displaystyle 5\)}%
\end{pgfscope}%
\begin{pgfscope}%
\pgftext[x=0.110488in,y=0.934919in,right,]{\rmfamily\fontsize{8.000000}{9.600000}\selectfont \(\displaystyle 10\)}%
\end{pgfscope}%
\begin{pgfscope}%
\pgftext[x=0.110488in,y=1.145826in,right,]{\rmfamily\fontsize{8.000000}{9.600000}\selectfont \(\displaystyle 15\)}%
\end{pgfscope}%
\begin{pgfscope}%
\pgftext[x=0.110488in,y=1.356733in,right,]{\rmfamily\fontsize{8.000000}{9.600000}\selectfont \(\displaystyle 20\)}%
\end{pgfscope}%
\begin{pgfscope}%
\pgftext[x=0.110488in,y=1.567640in,right,]{\rmfamily\fontsize{8.000000}{9.600000}\selectfont \(\displaystyle 25\)}%
\end{pgfscope}%
\begin{pgfscope}%
\pgftext[x=0.110488in,y=1.778547in,right,]{\rmfamily\fontsize{8.000000}{9.600000}\selectfont \(\displaystyle 30\)}%
\end{pgfscope}%
\begin{pgfscope}%
\pgftext[x=0.110488in,y=1.989455in,right,]{\rmfamily\fontsize{8.000000}{9.600000}\selectfont \(\displaystyle 35\)}%
\end{pgfscope}%
\begin{pgfscope}%
\pgfsetroundcap%
\pgfsetroundjoin%
\pgfsetlinewidth{1.003750pt}%
\definecolor{currentstroke}{rgb}{0.000000,0.000000,0.000000}%
\pgfsetstrokecolor{currentstroke}%
\pgfsetdash{}{0pt}%
\pgfpathmoveto{\pgfqpoint{1.343587in}{0.263840in}}%
\pgfpathlineto{\pgfqpoint{1.343587in}{0.189391in}}%
\pgfpathlineto{\pgfqpoint{0.320984in}{0.189391in}}%
\pgfpathlineto{\pgfqpoint{0.320984in}{0.263840in}}%
\pgfusepath{stroke}%
\end{pgfscope}%
\begin{pgfscope}%
\pgftext[x=0.785803in,y=0.070199in,,]{\rmfamily\fontsize{9.000000}{10.800000}\selectfont no shortlist}%
\end{pgfscope}%
\begin{pgfscope}%
\pgfsetroundcap%
\pgfsetroundjoin%
\pgfsetlinewidth{1.003750pt}%
\definecolor{currentstroke}{rgb}{0.000000,0.000000,0.000000}%
\pgfsetstrokecolor{currentstroke}%
\pgfsetdash{}{0pt}%
\pgfpathmoveto{\pgfqpoint{3.109900in}{0.263875in}}%
\pgfpathlineto{\pgfqpoint{3.109900in}{0.183194in}}%
\pgfpathlineto{\pgfqpoint{2.025322in}{0.183194in}}%
\pgfpathlineto{\pgfqpoint{2.025322in}{0.263875in}}%
\pgfusepath{stroke}%
\end{pgfscope}%
\begin{pgfscope}%
\pgftext[x=2.490141in,y=0.070199in,,]{\rmfamily\fontsize{9.000000}{10.800000}\selectfont with shortlist}%
\end{pgfscope}%
\end{pgfpicture}%
\makeatother%
\endgroup%
    \caption{Users were more satisfied with their choices.\label{fig:satisfaction}}
\end{figure}

Figure~\ref{fig:satisfaction} shows the results for the final survey question. As we can see, the majority (47 users for 78\%) prefers or strongly prefers the shortlist interface in terms of choice satisfaction ($p < 0.001$; binomial test). Ten users reported no difference in satisfaction with their choices, and three reported greater satisfaction without the shortlist. 
One may wonder whether the overall satisfaction as reported after the experiment corresponds to the average satisfactions that people reported after each session. The absolute scores people gave after each session are also inline with the overall satisfaction. The average satisfaction score of the 240 sessions that used shortlists was 4.29, which is statistically significantly higher than the score of sessions without shortlists, 4.15 ($p < 0.05$ under a random permutation test). 

In users comments and feedback, they identified the winnowing capability as one of the strengths of the shortlist interface. To quote a user's comment:
\begin{quoting}
Still, I can't help but feel more confident in the options I chose with the first interface [shortlist interface]. I couldn't even point out which ones here were selected in the first interface, but the process of filtering to my top 5 choices - and then to my single winner - in each round really made me confident that I wasn't losing track of a good movie in the shifting sands of my short-term memory.
\end{quoting}

Looking at other basic interaction measures, we can see that users are indeed making use of the shortlist as a tool for coming up with a final decision.
Recall that in the 240 sessions where the shortlist was available, users made use of the shortlist in 224 of these sessions. Furthermore, in those 224 sessions, the final choice came from the shortlist over 95\% of the time (215 out of 224 sessions) or 90\% of the time (215 out of 240) that the shortlist was available. That means that shortlists were in fact used as a tool to memorize and compare choices. We see this as further evidence that the task-specific support of the interface also enables people to make better decisions.


\subsection{People explore more with shortlists}
\begin{table}[tbp]
  \centering
    \begin{tabular}{rlrrr}
    \toprule
          &       & \multicolumn{2}{c}{block} &  \\
          \cmidrule{3-4}
          & \multicolumn{1}{c}{} & 1st block & 2nd block & average \\
          \midrule
    \multicolumn{1}{c}{\multirow{2}[3]{*}{\begin{sideways}interf.\end{sideways}}} & with shortlist & 211.7 & 135.4 & 173.6 \\
    \multicolumn{1}{c}{} & no shortlist & 144.0 & 90.8  & 117.4 \\[4pt]
     &  average     & 177.9 & 113.1 &  \\
    \bottomrule
    \end{tabular}
    \caption{Time-to-decision per session in seconds.\label{tbl:time}}
\end{table}

We saw in the previous section that users effectively and frequently adopted shortlists into their decision-making process, improving their satisfaction.
We now explore in more detail how people interacted with the system -- measured quantitatively by time-to-decision and in the number of items displayed. Our first key result is that people take longer to arrive at a decision with shortlists. Table~\ref{tbl:time} shows the time-to-decision per session in seconds under each condition. With shortlists, users take just under three minutes on average to decide, whereas without shortlists, they merely take two minutes (compare the rows in the rightmost column). 

As the study progresses, the amount of time a user spends per session may change for a number of reasons such as increasing familiarity with the task, fatigue, increasing comfort with the interface. We can thus compare average times also across users that experienced an interface first (``1st block'') versus the interfaces used later in the study (``2nd block'').  
Comparing values column-wise, we can see that the values in the first row are always larger than in the second row ($p < 0.01$; random permutation test with Bonferroni correction). We can also see that people take substantially less time in the second block (bottom average) regardless of interface. Possible reasons for this observation include learning effects or the fact that people are usually more tired in the second block. 
\begin{table}[htbp]
  \centering
    \begin{tabular}{rlrrr}
    \toprule
          &       & \multicolumn{2}{c}{block} &  \\
          \cmidrule{3-4}
          & \multicolumn{1}{c}{} & 1st block & 2nd block & average \\
          \midrule
    \multicolumn{1}{c}{\multirow{2}[3]{*}{\begin{sideways}interf.\end{sideways}}} & with shortlist & 155.1 & 111.0 & 133.0 \\
    \multicolumn{1}{c}{} & no shortlist & 102.4 & 76.6  & 89.5\\[4pt]
     &  average     & 128.7 & 93.8  &  \\
    \bottomrule
    \end{tabular}
  \caption{Unique items displayed per session.\label{tbl:displayed_items}}
\end{table}

To complement time-to-completion, we also report the unique number of displayed items as a measure of user engagement. The results for the same conditions as before are reported in Table~\ref{tbl:displayed_items}. As we can see, the same trends that we found for time-to-decision also hold up for the number of unique items displayed. With shortlists, users browse roughly 1.5 times as many items as without shortlists (rightmost column). Again, we see by comparing values column-wise that users always browsed through more items when given a shortlist than without it ($p < 0.01$; random permutation test with Bonferroni correction). Taken together, we saw that users not only take more time, but they also browse through more items when given a shortlist. In Table~\ref{tbl:implicit_feedback_overall} of Section~\ref{sec:implicit_feedback_overall} we will also see that this translates to an increased number of items examined. 
 
Another interesting observation is that users become more efficient over time. We can see from Table~\ref{tbl:time} that the average time-to-decision decreases by approximately 40\% when going to the second block, whereas the number of displayed items only falls off by about 30\%. This means that people spent less time per item in the second block, indicating they got more efficient. In summary, we saw that people explore more items and take longer to decide when given a shortlist. At the same time, however, they enjoy the experience more overall, as the previous subsections showed.  Thus, the users perceive the ability to explore more without the danger of forgetting as a strong positive even though they spend more time before making a decision.  Similar to an anytime algorithm, the shortlist enables a user to easily stop at any point and select from the list should they choose to stop exploring.

\subsection{People explore differently with shortlists}
We just saw that users explore more and longer with shortlists. We also saw that they were more satisfied with their choices.  However, users may take more time in a session because it takes them longer to find a selection they want or because they are taking time to build confidence that an item they have seen is actually what they would like to select. 
In this section, we answer this question by examining user behavior after the user's eventual chosen item was displayed for the first time in the session.  To this end, we consider the number of unique items displayed to the user after the eventually chosen item was first displayed and the relative position in the session where the chosen item was first displayed to the user. As defined in Section~\ref{sec:user-study}, {\em displayed} for an item means an item was visible on a user's viewport. The particular set of displayed items is therefore determined by how a user paged through results, applied facets to filter the set of movies, etc.


\begin{table}[htbp]
  \centering
    \begin{tabular}{rlrrr}
    \toprule
          &       & \multicolumn{2}{c}{block} &  \\
          \cmidrule{3-4}
          & \multicolumn{1}{c}{} & 1st block & 2nd block & average \\
          \midrule
    \multicolumn{1}{c}{\multirow{2}[3]{*}{\begin{sideways}interf.\end{sideways}}} & with shortlist & 115.65 & 65.96 & 90.81 \\
    \multicolumn{1}{c}{} & no shortlist & 41.54 & 37.89 & 39.72 \\[4pt]
     &  average     & 78.60 & 51.93 &  \\
    \bottomrule
    \end{tabular}
  \caption{Items displayed to the user after displaying the user's eventually chosen item for the first time.\label{tbl:further_back}}
\end{table}

Table~\ref{tbl:further_back} reports the number of unique items that were displayed after a user encountered the final chosen item in the session for the first time. Interestingly, we can see that with the shortlist, choices lie further back in the user's session history: the average number of unique items that were displayed after encountering the chosen item for the first time is more than doubled with the shortlist interface than the interface that provided no shortlist (compare rows in the rightmost column).

A simple explanation for this is that shortlists give the user the ability to easily get back to any item, even though it occurred far back in the past. However, it is also important to note that users with the shortlist actually choose to continue browsing after seeing a good item. This is in line with results in the next subsection, where we see that people adapt their decision-making strategies to the interface.

In order to normalize across session lengths, we also examine the relative position of chosen items in the list of displayed items. To do this, we order all displayed items of a session in the order they were displayed to the user. We only keep the position of the first occurrence of each item, so that if a user revisits items, the overall statistic remains stable. The relative position is now calculated as the position of the chosen item in the list of displayed items, divided by the total number of items displayed to the user in the session. In other words, we measure where in the session a user first encountered the item that she finally chose, normalizing for different session lengths. As an example, if a user picks an item that she saw in the very beginning of her session but continued exploring for a long time before making a decision, this would yield a relative position of almost zero. Likewise, if a user selects an item immediately after seeing it for the first time, we would see a value of close to one.

\begin{figure}[thb]
    \centering
    \input{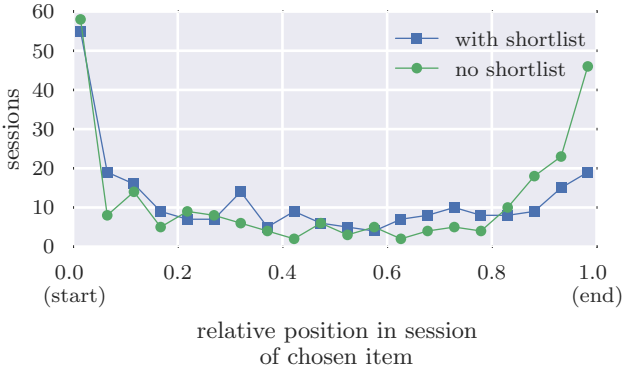}
    \caption{Without the shortlist, users are more likely to pick items towards the end of their session.\label{fig:relative_positions}}
\end{figure}

Figure~\ref{fig:relative_positions} shows the distribution of the relative positions of items selected for the two different interfaces. 
The first thing to notice is that in more than 50\% of sessions and with both interfaces, the chosen item was first displayed within the first 5\% of the session.
This is reasonable since the default ordering showed recent highly-rated movies first and users tend to prefer these. However, even though an interesting item was encountered early on, users continue to browse before settling on the decision.  Additionally, in both interfaces we see another increase in the number of sessions where items were picked at the very end. This is indicative of satisficing behavior (cf.\ Section \ref{sec:strategy}), where a user terminates the search immediately after encountering an item that meets a minimum personal threshold of quality \cite{simon1956rational}. 

Comparing user behavior under the two different interfaces, we make the following additional observations. First, there are slightly more sessions in which the final choice comes from the very beginning of the session in the no shortlist case. 
This makes sense given that with no shortlist interface, users had to keep track of options themselves, and humans remember items from the beginning or ends of lists better than from an intermediate position \cite{murdock1962serial}. 
Second, there is another, even more prominent difference toward the end of sessions: without shortlists, users are more likely to terminate a session shortly after encountering an interesting item.  
By comparing the heights of the rightmost points in Figure~\ref{fig:relative_positions}, we see that users in the no shortlist condition are twice as likely to show satisficing behavior than users in the shortlist condition -- stopping immediately after finding a minimum quality item.  There likely is an additional refinding effect at work here since refinding recently displayed items is much easier than refinding items displayed in the more distant past.  In contrast, when the shortlist interface is used, we see the ability to select items regardless of initial display point is much more evenly spread except for the spike at the beginning of the session where highly desirable items are likely to occur.

\subsection{Interface influences choice of strategy \label{sec:strategy}}
We just saw that people explore differently when they have the shortlist available. That raises the question of whether users always employ the same strategy and 
many of our users happened to have a strategy that is well-supported by a shortlist, or whether the design of the interface 
modifies the strategy choice of the users.
To get at this question, we asked after each block for users to self-report the strategy they used to make a choice. We offered them the following four options:
\begin{enumerate}[noitemsep]
\item I selected the first movie I thought was good.
\item I kept track of the single current best choice as I went along.
\item I kept track of candidate movies that were good and then selected one among them.
\item Other - please specify.
\end{enumerate}

The first option was targeted at users who psychologists refer to a satisficers \cite{simon1956rational}, i.e., these users just want to pick something that is good enough.  The second and third options aim more at utility maximizing users, i.e., users that want to find the best out of all available options.  However, the second option represents a strategy where users are willing to instantly determine whether an option surpasses all previously considered options while the third option is an explore-and-curate strategy where the user defers making a final decision among possible candidates.


\begin{table}[htb]
  \centering
    \begin{tabular}{lrp{0.0cm}rrp{0.0cm}r}
    \toprule
    \multicolumn{1}{r}{} & \multicolumn{3}{c}{shortlist last} & \multicolumn{3}{c}{shortlist first} \\
    \cmidrule(l{2pt}r{2pt}){2-4} \cmidrule(l{2pt}r{2pt}){5-7}
         strategy & no sl &\multicolumn{2}{r}{with sl} & with sl && no sl \\
    \midrule
    first good & 7 &$\rightarrow$     & 1    & 2  &$\rightarrow$    & 16\\
    single track & 5   &$\rightarrow$   & 4     & 3   &$\rightarrow$   & 7 \\
    multiple track & 15  &$\rightarrow$   & 24    & 19  &$\rightarrow$   & 3 \\
    other & 3 &$\rightarrow$     & 1     & 6   &$\rightarrow$   & 4 \\
    \midrule
    total & 30     && 30    & 30    && 30 \\
    \bottomrule
    \end{tabular}%
  \caption{People switch their strategies depending on the interface.\label{tbl:strategies}}
  \vspace*{-\baselineskip}
\end{table}%


The survey results are reported in Table~\ref{tbl:strategies}. The main insight from this table is that people do not seem to have a fixed strategy, but choose their strategy depending on the interface. Starting with the group of users that saw the shortlist interface last, we can see that 15 users were following an explore-and-curate approach in the no shortlist interface which they used first. However, when they move to the shortlist interface, more users (24) adopt the explore-and-curate approach ($p < 0.01$; binomial test). Further analysis of the transition matrix also confirmed that users with a satisficer approach of picking the first good item now switched to the explore-and-curate strategy (6 users). 

Even more interesting is what happens to users in the group that used the shortlist interface first. When these users move to the no shortlist interface, they seem to be upset and switch to a more greedy strategy of either tracking no item or just one. More specifically, while the shortlist was available, 19 users employed an explore-and-curate strategy and only two users followed a satisficer approach. Once the shortlist became unavailable, 16 users adopt the satisficer approach and the number of people with the explore-and-curate strategy reduced to three. 
This is consistent with our premise that users incur substantial cost when keeping items in short-term memory, and that this cost is reduced through the shortlist interface.

In summary, we saw that the availability of digital memory greatly influences the way people approach decisions. In particular, the question of whether someone shows maximizing or satisficing behavior is influenced by the cost of acquiring and storing information. 
Consistent with bounded rationality, when information can be stored easily, the decreased cognitive load enables the user to explore more items. 
The finding that users adapt their strategies to the environment has also been confirmed by other studies \cite{payne1988adaptive}.

\subsection{Discussion}
While there are many advantages to the controlled setting that we reported above, there are also some limitations that future work should try to address. For example, users selected movies but never actually watched them as that would extend the entire study containing multiple sessions over many hours or days. Another limitation is that we asked people to make eight choices in quick succession.  In practice, these choices would be spread out over time and we might find less of a drop-off in engagement with this more natural cadence. Hence, it would be interesting to connect our results with decision making in the wild -- where people will watch a movie after deciding on it.


\section{How Do Shortlists Impact Recommendation Quality?}
We have seen in the previous section that shortlists produce a better user outcome. That is, users prefer the shortlist interface, are more satisfied with the choices they make, and they also explore more. Now we turn to the question of whether shortlists also help improve recommendation quality. There are good reasons to believe so: we saw that users both interacted longer with the system and also browsed through more items when they had a shortlist. As the next subsection will show, people also generate more implicit feedback.  

Again, our goal is to study whether implicit feedback generated from shortlist sessions can be used to improve within-session recommendation. Ideally, we would like to have a recommender system in use in order to study interface effects in isolation. Since this is not the case, we follow a methodology similar to how recommender system studies are performed over logs of interaction data \cite{jannach2015shortterm}. In short, we will try to predict a single session's chosen item based on some subset of the user's actions in this session. 

\subsection{Do shortlists lead to more implicit feed-back?}
\label{sec:implicit_feedback_overall}
In this section, we compare the two different interfaces with respect to the amount of implicit feedback that they generate for learning. Following the common practice in recommendation, we look at direct interactions with an item as implicit expressions of user interest.  As described in Section \ref{sec:user-study} we defined ``examined'' items as those that a user has clicked on the item to get detailed information such as actor lists, synopses, etc., while ``shortlisted''items are those added to the shortlist.  It is possible to shortlist an item without examining the details of the item.  

\begin{table}[htb]
  \centering
    \begin{tabular}{lrrr}
    \toprule
          &       & \multicolumn{2}{c}{interface} \\
        \cmidrule{3-4}
          & type  & with sl & no sl  \\
    \midrule
    examined & all   & 4.43  & 3.04 \\
    examined & unique & 3.94  & 2.75 \\
    \bottomrule
    \end{tabular}
    \caption{The average number of items per session whose details were examined in each interface.\label{tbl:implicit_feedback_overall}}
\end{table}%

Table~\ref{tbl:implicit_feedback_overall} reports the average number of interactions per category (examined items, shortlisted items) per session. We report both the number of unique item interactions and the total number of interactions as indicated in the ``type" column.
The first thing to note is that users interact with more items when they have shortlist support, both in terms of unique items as well as the number of total items. With a shortlist, users examine over one item more on average per session (3.94 vs 2.75 items). Given the small number of examined items, this constitutes a relative increase of about 30\% in feedback data. 

\begin{table}[bht]
  \centering
    \begin{tabular}{lr}
    \toprule
      & with sl\\
    \midrule
    shortlisted or examined &  5.71   \\
    \hspace*{0.15in}examined &  3.94   \\
    \hspace*{0.3in}examined but not shortlisted &  2.24   \\
    \hspace*{0.3in}examined and shortlisted &  1.70   \\
    \hspace*{0.15in}shortlisted &  3.48   \\
    \hspace*{0.3in}shortlisted but not examined &  1.78   \\
    \hspace*{0.3in}examined and shortlisted &  1.70   \\
    \bottomrule
    \end{tabular}
    \caption{The average number of unique items per session with interactions of examined or shortlisted when using the shortlist interface.\label{tbl:implicit_feedback_shortlist}}
\end{table}%

In Table~\ref{tbl:implicit_feedback_shortlist}, we break down interaction types within the shortlist interface.  In particular, we see that shortlists provide us with a second type of implicit feedback. In the shortlist interface, we additionally get to observe clicks that reflect adds to the shortlist. The indentation in the table helps visualize the subset relationships; that is ``shortlisted {\em and} examined'' is a subset of the items that are ``examined'' which is a subset of those that are ``shortlisted {\em or} examined''.  Thus this is a table version of the values within each cell of the Venn diagram for these two types of interactions and ``examined and shortlisted'' is listed twice to reflect the subset relationships.

From the table, we can see that people use the shortlisting mechanism quite extensively with 3.48 items per session. This raises the question of how much overlap there is between examined and shortlisted items or whether shortlisted items yield additional information to knowing the set of examined items. Interestingly, the set of shortlisted items and the set of examined items only overlap partially; their intersection contains about 1.7 items on average. There are more items that were examined but not shortlisted (2.24) than items that were shortlisted but not examined (1.78). Presumably movies that were added to the shortlist without examination are movies the user was aware of prior to the study.  By having the shortlist, we gain a stronger signal of a user's preferences with respect to already known items even when an item is not eventually chosen -- in the absence of a shortlist this type of implicit feedback may be difficult to ascertain outside of eye-tracking.

Furthermore, by comparing the number examined (3.94) with the number examined but not shortlisted (2.24), we can see that more than half of examined items do not end up in the shortlist. This means that shortlists actually are used as a curating mechanism since not all examined items also get shortlisted. Moreover, while both may be signals of a user's interest, it implies that the feedback signals (examined vs.\ shortlisted) may also be different in their nature. Overall, we can see in the last row that by considering both shortlisted as well as examined items, we have an average of 5.71 items as interaction feedback data. Recall from Table~\ref{tbl:implicit_feedback_overall} that in no shortlist sessions we only had an average of 2.75 items with interaction feedback, that means that shortlists were able to approximately double the amount of data we obtained from user interactions.

In summary, we have seen that users give up to two times as much \emph{more feedback} with shortlists, and the kind of feedback we obtain from considering add-to-shortlist interactions is also \emph{different}. 

\subsection{Does the increased feedback quantity improve recommendation quality?}
In the previous section, we demonstrated that the shortlist interface leads to interaction feedback on approximately two times as many unique items as in the no shortlist condition. This raises two interesting questions that we will investigate in this and the following subsection. First, there is the question of whether the increase in the amount of feedback data also translates to an increase in recommendation quality. Second, we want to know whether distinguishing the types of feedback (examined or shortlisted) is essential for learning.

We start by describing the overall recommendation experimental setup that is used in this and the following subsection. Since we assume a session-based setting, we need to train and test on the same session. Our basic setup is similar to \cite{jannach2015shortterm}. The basic idea is to train on implicit feedback from the session, and then use the model trained on feedback data to predict the heldout final chosen item of a session from a random subset of movies. 

The overall protocol is as follows. Each session, $S \in \mathcal{S}$, forms one prediction problem, for which we train a ranking SVM \cite{joachims2005support} where we aim to predict the user's final selection based on a sample of observed interaction data. We split the set of sessions $\mathcal{S}$ randomly into a set of evaluation sessions $\mathcal{S}_{eval}$ and validation sessions $\mathcal{S}_{val}$ in a 3:1 ratio. We ensure these proportions on a user level, so if a user had four sessions, three would go into the evaluation set and one into the validation set.   We use the validation data to tune the ranking SVM's hyperparameter $\lambda$ by measuring performance for various hyperparameter choices $\{\lambda\}$ and selecting the best on the validation set $\mathcal{S}_{val}$.  We then use this value for the hyperparameter when training a model for each of the sessions in the evaluation set, $\mathcal{S}_{eval}$.   Note that since we learn a separate model for each session, there are no parameters beyond the hyperparameter shared across users. 

For each session in the evaluation set, we must further divide the data into data that we can use for training the session model and data that we can use for testing that model's generalization with respect to the session.
First, we constructed feature vectors for each movie. For the features, we considered a broad range of properties from OMDb, including a movie's year, popularity, actors, directors and tf-idf vectors of the plot synopsis. Then, for a session, $S$, let $V_S$ be the items that were displayed in the session, and $A_S$ all items that were in the inventory, i.e., all 1030 movies. Furthermore, let $x_S^* \in V_S$ be the movie that was the chosen item in a session.  Ideally a model that generalizes well will be able to rank the user's chosen item,  $x_S^*$, above alternatives.  To sample from the session data, the detailed process was as follows:
\begin{enumerate}
\item Create a test set $D_{\mathit{test},S}$ by randomly sampling 99 movies from $A_{S} \setminus \{x_S^*\}$ and adding $x_S^*$ to this subset.
\item Train on $D_{\mathit{train},S} = V_{S} \setminus D_{\mathit{test},S}$, i.e. all items displayed to the user that were not held out for the test set, $D_{\mathit{test},S}$.  To define the target input ranking $D_{\mathit{train},S}$, we used the following rule to interpret the implicit feedback: $\{shortlisted, examined\} \succ displayed$, i.e., all the items that got shortlisted or examined on had to be ranked before items that only were displayed. This is based on the common assumption that users reveal their preferences through clicks. 
\item Test on $D_{\mathit{test},S}$, where $x_S^*$ is ranked at position 1 is ideal. Measure the reciprocal rank (RR) of $x_S^*$ in the predictions on $D_{test}$. Ranking the chosen item on top of all other options yields a RR value of 1, whereas ranking it last would result in a RR value of 1/100.
\end{enumerate}


Recall that the question that we want to answer in this subsection is whether the additional implicit feedback data obtained with the shortlist is non-redundant and can thus help improve recommendation performance.
To answer this question, we measure recommender performance for our system above trained either on:
\begin{enumerate}
\item[(i)] $\mathcal{S}^{with\_sl}$ defined as all sessions that had the shortlist (240 in total); or
\item[(ii)] $\mathcal{S}^{no\_sl}$ defined as all the session where users had no shortlist (also 240 in total).
\end{enumerate}
Note that each set of sessions, $\mathcal{S}^{C}$, for a condition, $C\in \{\mathit{with\_sl, \ no\_sl}\}$, was partitioned as described earlier into $\mathcal{S}^{with\_sl}_{eval}$ (180 sessions) for training and $\mathcal{S}^{with\_sl}_{val}$ (60 sessions) for validation.  Because of the balance in our user study, each user is equally represented in both conditions, thus the difficulty of both tasks should be comparable with respect to predicting each user's preferences.

The results are listed in Table~\ref{tbl:shortlist_recommendation}. We can see that 
with feedback from shortlist sessions, our recommendation performance (measured as the mean reciprocal rank) is almost twice as good as with feedback from sessions where no shortlists were available. This difference is also statistically significant ($p < 0.001$ under a random permutation test with $n=10^6$ samples). Note also that both systems perform better than a random ranker, which puts the picked item in position $i$ with uniform probability. 

\begin{table}[t]
  \centering
    \begin{tabular}{rr}
    \toprule
          & MRR \\
    \midrule
    $\mathcal{S}^{with\_sl}_{eval}$ & 0.11875 \\[3pt]
    $\mathcal{S}^{no\_sl}_{eval}$ & 0.06325 \\[3pt]
    random & 0.05200 \\
    \bottomrule
    \end{tabular}
  \caption{Using feedback from shortlist sessions improves recommendation quality\label{tbl:shortlist_recommendation}}
\end{table}

\subsection{Does modeling extra granularity of the\\feedback help?}
As we have seen previously, there exists a substantial number of items that only get examined but never added to the shortlist. In these cases, we might infer that the user actually liked these items less than the ones he both examined and shortlisted. The question we address in this section asks whether we should actually distinguish such cases during training or whether we can conflate examined and shortlisted items since both may represent a user's general preferences as they generalize to predicting the final chosen item.
We now keep the sessions we train on fixed. We always use $\mathcal{S}^{with\_sl}$, i.e., only sessions that had the shortlist, but vary the way in which we construct the training rankings. Our models use either:
\begin{enumerate}
\item[(i)] $\mathcal{D}^{coarse}_{train}$ which uses the same preferences as before, i.e., $\{shortlisted, examined\} \succ displayed$ or
\item[(ii)] $\mathcal{D}^{fine}_{train}$ in which we prefer shortlisted items over everything else, i.e., $shortlisted \succ examined \succ displayed$. Note that examined here refers to the items that got examined but not shortlisted. 
\end{enumerate}

As we can see from the results in Table~\ref{tbl:granularity}, there is virtually no performance difference between the two models. The difference is also statistically not significant, meaning that it did not pay off to distinguish between the various types of feedback. Note also that the first line repeats the same result as in Table~\ref{tbl:shortlist_recommendation}. Even though intuitively, items that get shortlisted may carry more meaning to the user, it might be the case that for small amounts of examples this distinction does not help better fit the model or that the user decided not to shortlist some items for other reasons not indicative of overall interest (e.g. upon reading the description the user remembers having already seen the movie). 
\begin{table}[t]
  \centering
    \begin{tabular}{rr}
    \toprule
          & MRR \\
    \midrule
    $\mathcal{S}^{with\_sl}_{eval}$ and $\mathcal{D}^{coarse}_{train}$ & 0.11875 \\[3pt]
    $\mathcal{S}^{with\_sl}_{eval}$ and $\mathcal{D}^{fine}_{train}$ & 0.11585 \\[3pt]
    random & 0.05200 \\
    \bottomrule
    \end{tabular}
  \caption{Distinguishing between different levels of feedback does not further improve recommendation performance. \label{tbl:granularity}}
  \vspace{0pt}
\end{table}

\subsection{Discussion}
In general, we see the results of the experiment as evidence for preferring user engagement over feedback discrimination; i.e., when designing interfaces, it pays off to think more about encouraging user engagement rather than discriminating different qualities of implicit feedback. It would be interesting for future research to investigate this question in more detail, i.e., answering the question when exactly different quality levels of implicit feedback could be of use.

\section{Related Work}
Our work in this paper is located in the intersection of human-computer interaction, psychology and machine learning. Each aspect of shortlists draws ideas from a different area. The usability aspect of shortlists is most strongly related to HCI, and we showed that user satisfaction under the shortlist interface is improved. Research in cognitive psychology helps us understand why having an external memory aid supports decision making. Lastly, from a machine learning perspective, shortlists are important as means to obtain more implicit feedback for learning. We will now discuss each of the related areas in more depth.

Starting with HCI, there are a number of systems that were designed with the goal of aiding people in decision making. As there is a large body of work on general decision support systems \cite{power2002decision}, we only discuss work pertaining to search and recommendation \cite{chen2013human}. Ruetsalo {\em et al.}\ \cite{ruotsalo2013supporting} propose a system for information retrieval tasks where a user model gets adapted during the search process, allowing the user to update feature weights after each query. We, in contrast, do not ask for explicit feedback in any form, but assume users are rational enough to only shortlist items that have relatively high utility. Also in information retrieval, Jia and Niu \cite{jia2014should} propose an interface that helps people know when to stop exploring. Drucker {\em et al.}\ \cite{drucker2005visual} present a visual way of supporting movie selection in groups -- an interesting scenario we would like to like to study in the future. In contrast to their work, we propose shortlists not as an entire system to solve an end-to-end task, but rather as a component that provides digital memory. Hence, our approach can be seen as complementary to these systems -- one can imagine adding shortlists to them as an additional component. 

From research in psychology, we know that there are clear limits on people's short term memories. Numbers range from three up to nine chunks of information that could be held in memory at the same time \cite{crowder1976principles, baddeley1999essentials}. Not only is the amount of information in short-term memory limited, but it also decays fairly quickly if not used; decay times around 18 seconds have been reported in studies \cite{revlin2012cognition}. Shortlists are fighting these two limitations in parallel: users can both remember more items and recall them at any time of the session. The role of memory in recommender systems as well as the need for support for it is also further discussed by Del Missier \cite{delmemory}. There is also a large body of research on decision making in psychology. Jameson \cite{Jameson14DMRS} summarizes support principles for decision making in the context of recommendation in a high-level framework called ARCADE. In this framework, shortlists can be seen as realizations of two strategies. Namely, shortlists may help \emph{advise} the user in making better decisions by suggesting a winnowing approach to choice making, and shortlists can \emph{represent} the current set of candidates a user is considering. In summary, we saw that shortlists have well motivated cognitive and psychological foundations.

A key concern in machine learning is the availability of labeled data, often obtained in the form of human feedback. Many machine learning approaches assume \emph{explicit} feedback from the user. Explicit feedback means that users are explicitly asked to provide some form of feedback on the output produced by a ML system \cite{kelly2003implicit}. This is different from \emph{implicit} feedback that is obtained as a by-product of the user interacting with a system. Schemes for explicit feedback elicitation thus are all of an invasive nature, ranging from minimally to strongly invasive. On the minimal end of the spectrum, there are systems in information retrieval \cite{salton1971smart,Salton90improvingretrieval} or recommendation \cite{dooms2011online} that imagine that users provide optional feedback on items using a thumbs up or down mechanism. More invasive is active learning \cite{settles2009active}, where users are iteratively queried for more feedback and the system also decides which options a user needs to give feedback on. 
The main limitation of explicit feedback is that users are reluctant to give it since it provides no immediate benefit to them; participation rates of under 1\% were found in practice \cite{dooms2011online}.

Implicit feedback overcomes data scarcity by relying on user actions. It assumes users reveal their interests through the actions they take. Several methods try to harness this fact. For example, in information retrieval evaluation, implicit feedback is used to infer preferences for rankings \cite{joachims2002interleaving,radlinski2008interleaving}. 
A line of work called gamification \cite{hamari2014does} tries to set up \emph{external} rewards (e.g., badges, leaderboards, etc.) so that users will change their behavior accordingly. The interface that we introduce in this paper takes a different approach: we leverage the \emph{internal} motivation of users to make good choices, and users receive immediate benefits (decreased cognitive load, greater satisfaction) when using the shortlist.

Lastly, there is a growing interest in recommendation and decision making on a session-based level. Cremonisi \emph{et al.}\ \cite{cremonesi2012decision} study decision making in recommendation systems for hotel search. The authors perform a user study where decision making happens either with the help of a recommender system or without. Interestingly, they found that the number of examined items as well as the time-to-decision increased when users employed a recommender system. This again shows the need to consider both interface design and feedback elicitation at the same time.  On the algorithmic side, several approaches are designed to learn from session-based data \cite{mobasher2002using,ricci2003product,jannach2015shortterm}. The work done by Jannach \emph{et al.}\ \cite{jannach2015shortterm} also adopts a session-based approach to recommendation, but, in contrast to us, assumes that a long term interest profile is available. A challenge that we had to face was also to learn from multiple levels of implicit feedback. Most work assumes that one has access to enough users so that graded relevance labels can just be integrated into standard collaborative filtering models \cite{shi2013graded, lerche2014graded}. Our approach did not assume that feedback was available from other users since we adopted a cold-start session-based scenario.

To the best of our knowledge, this is the first work studying shortlists as design patterns to improve both user satisfaction and feedback elicitation. We demonstrated empirically that users generate more implicit feedback, and that this feedback in turn can be used to improve recommendation quality. Shortlists can be also seen as a bridge between the diverging goals of end users and system designers.

\section{Future Work}
Although we only focused on movie recommendation, we believe that the concept of shortlists or even broader, digital memory, can be applied to a more general class of tasks, such as trip planning or online shopping. 
In these scenarios, it might be even more important to obtain more task-specific feedback since item inventories are changing constantly, and long-term preferences might not be sufficient. The idea of digital memory is also backed up by research in cognitive psychology, and we confirmed its effectiveness in our experiments. Hence, digital memory is a valuable asset for interface design since it eases cognitive burden and incentives and engages users. Interestingly, it was powerful enough to even change self-reported user behavior. Instead of satisficing at the first minimally good item, many users adopted an explore-and-curate strategy under the shortlist interface. This evidence suggests that people's effort and task involvement is strongly coupled to the interface given -- factors that are highly relevant for e-commerce applications. 

Future work could study the use of digital memory in other scenarios that differ from the movie domain considered here -- both in terms of domain knowledge and investment. For example, shortlists can be valuable in domains where the options are completely unknown to the user (e.g., restaurants in a new city) since in these scenarios, there would be more emphasis on exploration. It would also be interesting to look at shortlist usage in domains where decisions involve a larger risk, e.g., laptop shopping or job search. 

Other interesting directions are the interplay of short-term interests, as reflected in shortlists, and long-term interests, for example given by a wish list. There is also the possibility of doing recommendation based on the entire content of shortlists, similar to next-basket recommendation for e-commerce websites \cite{rendle2010factorizing}. A possible use case of this would be to prepopulate a user's shortlist. 

\section{Conclusions}
We demonstrated the importance of designing recommender systems holistically in our user study. Introducing shortlists yielded both improvements in user satisfaction and downstream recommendation performance. In particular, we saw that users preferred an interface with shortlist support, they were more satisfied with their choices and stay engaged longer when they had shortlist support. This engagement resulted in additional implicit feedback that improved the quality of recommendations by nearly a factor of two.

This research was funded in part through NSF Awards IIS-1247637, IIS-1217686, and IIS-1513692.

\enlargethispage{-\baselineskip}
\bibliographystyle{abbrv}

\begin{thebibliography}{10}

\bibitem{baddeley1999essentials}
A.~D. Baddeley.
\newblock {\em Essentials of human memory}.
\newblock Psychology Press, 1999.

\bibitem{chen2013human}
L.~Chen, M.~de~Gemmis, A.~Felfernig, P.~Lops, F.~Ricci, and G.~Semeraro.
\newblock Human decision making and recommender systems.
\newblock {\em TiiS}, 3(3):17, 2013.

\bibitem{cremonesi2012decision}
P.~Cremonesi, A.~Donatacci, F.~Garzotto, and R.~Turrin.
\newblock Decision-making in recommender systems: The role of user's goals and
  bounded resources.
\newblock In {\em RecSys: Workshop on Human Decision Making in Recommender
  Systems}.

\bibitem{crowder1976principles}
R.~G. Crowder.
\newblock {\em Principles of learning and memory.}
\newblock Lawrence Erlbaum, 1976.

\bibitem{delmemory}
F.~Del~Missier.
\newblock Memory and decision making: From basic cognitive research to design
  issues.
\newblock 2014.

\bibitem{dooms2011online}
S.~Dooms, T.~De~Pessemier, and L.~Martens.
\newblock An online evaluation of explicit feedback mechanisms for recommender
  systems.
\newblock In {\em WEBIST}, pages 391--394, 2011.

\bibitem{drucker2005visual}
S.~M. Drucker, T.~Regan, A.~Roseway, and M.~Lofstrom.
\newblock The visual decision maker--a recommendation system for collocated
  users.
\newblock In {\em ACM SIGDUX}, 2005.

\bibitem{hamari2014does}
J.~Hamari, J.~Koivisto, and H.~Sarsa.
\newblock Does gamification work? -- {A} literature review of empirical studies
  on gamification.
\newblock In {\em HICSS}, pages 3025--3034, 2014.

\bibitem{Jameson14DMRS}
A.~{Jameson}.
\newblock Recommender systems as part of a choice architecture for {HCI}.
\newblock In {\em {I}nternational {W}orkshop on {D}ecision {M}aking and
  {R}ecommender {S}ystems ({DMRS})}, 2014.

\bibitem{jannach2015shortterm}
D.~Jannach, L.~Lerche, and M.~Jugovac.
\newblock Adaptation and evaluation of recommendations for short-term shopping
  goals.
\newblock In {\em RecSys}, pages 211--218, 2015.

\bibitem{jia2014should}
Y.~Jia and X.~Niu.
\newblock Should {I} stay or should {I} go: Two features to help people stop an
  exploratory search wisely.
\newblock In {\em CHI '14 Extended Abstracts on Human Factors in Computing
  Systems}, pages 1357--1362, 2014.

\bibitem{joachims2002interleaving}
T.~Joachims.
\newblock Optimizing search engines using clickthrough data.
\newblock In {\em KDD}, pages 133--142, 2002.

\bibitem{joachims2005support}
T.~Joachims.
\newblock A support vector method for multivariate performance measures.
\newblock In {\em ICML}, pages 377--384, 2005.

\bibitem{kelly2003implicit}
D.~Kelly and J.~Teevan.
\newblock Implicit feedback for inferring user preference: a bibliography.
\newblock In {\em SIGIR Forum}, volume~37, pages 18--28, 2003.

\bibitem{lerche2014graded}
L.~Lerche and D.~Jannach.
\newblock Using graded implicit feedback for bayesian personalized ranking.
\newblock In {\em RecSys}, pages 353--356, 2014.


\bibitem{lidwell2010universal}
W.~Lidwell, K.~Holden, and J.~Butler.
\newblock {\em Universal principles of design, revised and updated: 125 ways to
  enhance usability, influence perception, increase appeal, make better design
  decisions, and teach through design}.
\newblock Rockport Publishers, 2010.

\balancecolumns

\bibitem{miller1956magical}
G.~A. Miller.
\newblock The magical number seven, plus or minus two: some limits on our
  capacity for processing information.
\newblock {\em Psychological Review}, 63(2):81, 1956.


\bibitem{mobasher2002using}
B.~Mobasher, H.~Dai, T.~Luo, and M.~Nakagawa.
\newblock Using sequential and non-sequential patterns in predictive web usage
  mining tasks.
\newblock In {\em ICDM}, pages 669--672, 2002.

\bibitem{murdock1962serial}
B.~B. Murdock~Jr.
\newblock The serial position effect of free recall.
\newblock {\em Journal of Experimental Psychology}, 64(5):482, 1962.

\bibitem{payne1988adaptive}
J.~W. Payne, J.~R. Bettman, and E.~J. Johnson.
\newblock Adaptive strategy selection in decision making.
\newblock {\em Journal of Experimental Psychology: Learning, Memory, and
  Cognition}, 14(3):534, 1988.

\bibitem{power2002decision}
D.~J. Power, R.~Sharda, and F.~Burstein.
\newblock {\em Decision support systems}.
\newblock Quorum Books, 2002.

\bibitem{radlinski2008interleaving}
F.~Radlinski, M.~Kurup, and T.~Joachims.
\newblock How does clickthrough data reflect retrieval quality?
\newblock In {\em CIKM}, 2008.

\bibitem{rendle2010factorizing}
S.~Rendle, C.~Freudenthaler, and L.~Schmidt-Thieme.
\newblock Factorizing personalized {M}arkov chains for next-basket
  recommendation.
\newblock In {\em WWW}, pages 811--820, 2010.

\bibitem{revlin2012cognition}
R.~Revlin.
\newblock {\em Cognition: Theory and practice}.
\newblock Palgrave Macmillan, 2012.

\bibitem{ricci2003product}
F.~Ricci, A.~Venturini, D.~Cavada, N.~Mirzadeh, D.~Blaas, and M.~Nones.
\newblock Product recommendation with interactive query management and twofold
  similarity.
\newblock In {\em Case-Based Reasoning Research and Development}, pages
  479--493. Springer, 2003.

\bibitem{ruotsalo2013supporting}
T.~Ruotsalo, K.~Athukorala, D.~G{\l}owacka, K.~Konyushkova, A.~Oulasvirta,
  S.~Kaipiainen, S.~Kaski, and G.~Jacucci.
\newblock Supporting exploratory search tasks with interactive user modeling.
\newblock {\em Proceedings of the American Society for Information Science and
  Technology}, 50(1):1--10, 2013.

\bibitem{salton1971smart}
G.~Salton.
\newblock {\em The SMART Retrieval System---Experiments in Automatic Document
  Processing}.
\newblock Prentice-Hall, Inc., 1971.

\bibitem{Salton90improvingretrieval}
G.~Salton and C.~Buckley.
\newblock Improving retrieval performance by relevance feedback.
\newblock {\em Journal of the American Society for Information Science},
  41:288--297, 1990.

\bibitem{settles2009active}
B.~Settles.
\newblock Active learning literature survey.
\newblock Technical Report 1648, University of Wisconsin, Madison, 2009.

\bibitem{shi2013graded}
Y.~Shi, A.~Karatzoglou, L.~Baltrunas, M.~Larson, and A.~Hanjalic.
\newblock {xCLiMF}: Optimizing expected reciprocal rank for data with multiple
  levels of relevance.
\newblock In {\em RecSys}, pages 431--434, 2013.

\bibitem{simon1955behavioral}
H.~A. Simon.
\newblock A behavioral model of rational choice.
\newblock {\em Quarterly Journal of Economics}, pages 99--118, 1955.

\bibitem{simon1956rational}
H.~A. Simon.
\newblock Rational choice and the structure of the environment.
\newblock {\em Psychological Review}, 63(2):129, 1956.

\end{thebibliography}

%
%

\end{document}